\documentclass[journal]{IEEEtran}
\usepackage{blindtext, graphicx}
\usepackage{mathrsfs}
\usepackage{cite}

\ifCLASSINFOpdf
\else
\fi
\usepackage{amsmath}
\usepackage{algorithm}
\usepackage{algorithmic}
\usepackage[caption=false,font=footnotesize]{subfig}
\usepackage{url}

\usepackage[usenames,dvipsnames]{color}
\usepackage{soul}

\begin{document}

\title{State Estimation in Power Distribution Systems Based on Ensemble Kalman Filtering}
\author{C\^ome~Carquex,
        Catherine~Rosenberg
        and~Kankar~Bhattacharya
\thanks{C. Carquex, C. Rosenberg and K. Bhattacharya are with the Department
of Electrical and Computer Engineering,  University of Waterloo, Waterloo, ON N2L 3G1, Canada, e-mail: \{cacarque, cath, kankar\}@uwaterloo.ca}
}


%

\markboth{}%
{}
%
\maketitle

\begin{abstract}
State estimation in power distribution systems is a key component for increased reliability and optimal system performance. Well understood in transmission systems, state estimation is now an area of active research in distribution networks. While several snapshot-based approaches have been used to solve this problem, few solutions have been proposed in a dynamic framework. In this paper, a Past-Aware State Estimation (PASE) method is proposed for distribution systems that takes previous estimates into account to improve the accuracy of the current one, using an Ensemble Kalman Filter. Fewer phasor measurements units (PMU) are needed to achieve the same estimation error target than snapshot-based methods. Contrary to current methods, the proposed solution does not embed power flow equations into the estimator. A theoretical formulation is presented to compute a priori the advantages of the proposed method vis-a-vis the state-of-the-art. The proposed approach is validated considering the 33-bus distribution system and using power consumption traces from real households.

\end{abstract}



\begin{IEEEkeywords}

Distribution system, distribution system state estimation, ensemble Kalman filter, phasor measurement unit, state estimation

\end{IEEEkeywords}

\IEEEpeerreviewmaketitle

\vspace{-6pt}
\section{Introduction}

\IEEEPARstart{T}{raditionally}, electric power distribution systems have been designed and operated as passive systems to meet the customers' demand. However, with transformation of the grid to a smart grid, the reliability and operational challenges of distribution systems have increased. An operator will need to manage the distribution system more closely in the future, requiring improved visibility of its states~\cite{paudyal_optimal_2011} 
which will involve real-time monitoring~\cite{ardakanian_real-time_2014}. Indeed, most solutions to smart grid related challenges at the distribution level assume a knowledge of the states of the system, and therefore essentially rely on Distribution System State Estimation (DSSE), which is a key function of supervisory control that some utilities have already began rolling-out \cite{atanackovic_deployment_2013}.
The state of a power system can be completely defined from the knowledge of all bus voltage magnitudes and angles at time $t$ \cite{monticelli_state_1999}; typically, state estimation is carried out based on measurements of variables such as the voltage magnitudes and angles, available from Phasor Measurement Units (PMUs).

State estimation of power systems is a well understood problem at the transmission level and is traditionally solved using a snapshot-based weighted least square (WLS) method which relies on high quality measurement data from PMUs \cite{monticelli_state_1999}. 
However, transmission systems generally have a limited number of buses and are equipped with many measurement devices since it is important to precisely monitor and control the system at all times. On the other hand, distribution systems comprise a large number of buses with little to no measurements available. 
While several recent studies have focused on developing low-cost, easy to deploy PMUs \cite{von_meier_micro-synchrophasors_2014, rodrigues_low_2016}, it is not practical to install PMUs at every distribution bus. If PMUs were to be placed at selected buses only, there would be infinitely many solutions to the DSSE problem. In order to reduce the number of possible solutions, pseudo-measurements can be used \cite{fantin_using_2014}, which are load forecasts computed ahead of time to aid DSSE in finding a ``good'' solution. Typically, a pseudo-measurement at a given load bus comprises an estimate of the expected active and reactive power consumptions at the bus. Load forecasting at the distribution level is difficult, hence pseudo-measurements are usually of poor quality. These fundamental differences, and the need for affordable solutions, mean that new state estimation approaches are needed for distribution systems.


Many studies have extended the WLS approach from transmission to distribution systems. A review of literature on the different state estimation techniques and their application to DSSE problems is presented in \cite{primadianto_review_2016}. One of the first applications of the snapshot approach to the DSSE problem was reported in \cite{ghosh_distribution_1997}, where a probabilistic formulation based on pseudo-measurements was used.
In \cite{schenato_bayesian_2014}, the power-flow equations were linearized and a computationally friendly solution method was proposed. The authors also showed that PMUs are needed for accurate state estimation. Compressed sensing theory was used for state estimation with sparse measurements in \cite{alam_distribution_2014}, while \cite{wang_revised_2004} used line-current magnitudes and angles. Finally a semi-definite programming approach was used to solve the DSSE problem in \cite{klauber_distribution_2015}. 

Several researchers have used Kalman filters in state estimation problems for transmission systems \cite{filho_forecasting-aided_2009}. 
However, in distribution systems, the poor quality of the pseudo-measurements renders such methods ineffective. Therefore, very few Kalman filtering based methods have been developed for DSSE and none improve over the WLS. Huang \textit{et al}. compared the extended Kalman filter to the unscented Kalman filter in \cite{huang_evaluation_2015}. From the reported results it was noted that there was no visible improvement in performance of the Kalman filter based methods over WLS. In \cite{sarri_state_2012} the impact of choice of the model and measurement covariance matrix on the performance of the extended Kalman filter was examined. It was noted from the results that the proposed filtering approach did not result in any performance improvement. 
The above discussed Kalman filter based approaches apply the methods directly from the transmission to distribution systems. The problem of poor quality of pseudo-measurements is alleviated by assuming that measurements are available at every bus in real-time or quasi-real-time, usually from synchronized smart-meters, which is not realistic. 


In this paper, a past-aware method for DSSE, named PASE (Past-Aware State Estimation), where the estimate at time $t$ depends on anterior estimates and based on the Ensemble Kalman Filter (EnKF) \cite{evensen_ensemble_2003} is presented. Applying the EnKF to this problem is non-trivial, since measurements from sources with different time-scales must be merged. Contrary to WLS and other approaches using different variations of the Kalman filter, PASE does not embed the power flow equations into the estimator, making it a versatile technique. Instead it relies on an external power-flow solver, which is left to the choice of the operator.

In a snapshot-based context where the state at time $t$ is computed independently of the estimates at times anterior to $t$, the WLS objective function provides the best performance possible (excluding ill-conditioned cases)\cite{singh_choice_2009}. Such an estimator is referred to as the State of the Art (SoA) in this paper, for the purpose of comparison. 

Specifically, the contributions of the work are threefold: 1) A maiden attempt is made to apply EnKF to a distribution system sparsely monitored by PMUs for state estimation; 2) An analytical framework is developed to evaluate the performance of PASE; 3) The theoretical results are validated via extensive simulations on a 33-bus distribution system and using power consumption traces from real households. The performances of the proposed PASE approach and WLS are compared and engineering insights are provided to understand the impact of each decision variable on the performance of PASE, as well as the trade-offs to make.
Based on the above discussions, the main message of this work is that PASE is the first technique to improve upon the SoA. It does so significantly when the elapsed time between two consecutive state computations is small (less than 15 minutes)
, i.e., less PMUs are needed to achieve the same estimation error.


The rest of the paper is organized as follows. 
The background and assumptions are presented in Section \ref{system_model}. The SoA method is presented in Section \ref{SoA} and the proposed PASE solution in Section \ref{proposed_solution}. The validation results are reported in Section \ref{validation}. Finally, the conclusions are drawn in Section \ref{conclusion}.

\vspace{-6pt}
\section{System and Assumptions}
    \label{system_model}
    
The assumptions are stated in this section. A three-phase balanced distribution system under normal operations is considered. The DSSE problem is solved by the local distribution company (LDC) using an appropriate computational platform. The following information is needed to implement DSSE, both with the SoA method and the proposed PASE method. 


\textbf{Computational timescale}: a new state estimate is computed every $\Delta T$. Typically in transmission systems, a time-step of 5 to 15 min is considered. In distribution systems smaller time-steps are needed because of higher load volatility, which can arise for example with high penetration of renewables. The value of $\Delta T$ has an impact on the computational burden. In this work time-steps from 6 seconds to 15 minutes are considered. Altogether, the choice of an appropriate timescale for DSSE problems is still an open question.


\textbf{Topology:} the distribution system has a radial topology and is defined by a set of buses $I$ of cardinality $|I|$ as well as a set of branches $B$ of constant and known impedances, connecting the buses. The substation transformer is modeled as a reference voltage source of magnitude $V_0$. 


\textbf{Measurements:} the subset $S \subseteq I$ of buses are equipped with PMUs that monitor every $\Delta T$ both the bus voltage magnitudes ($V_s$) and bus angles ($\delta_s$). The measurements reported by the PMUs are assumed to be unbiased and the variance of the error of the readings is known. These assumptions are commonly made in state estimation works \cite{monticelli_state_1999}. A broadband communication infrastructure is available to transmit the measurements with low latency and high reliability. The PMUs are placed in the distribution system according to a given mapping $\mathscr{S}$.


\textbf{Pseudo-measurements:} these are forecasts that ``measure'' both active and reactive powers. They are available for each bus $i$ in $I$. 
Forecasts are made at periodic intervals $\Delta T'$, typically once a day for the next day (day-ahead forecast). At the time of computation, the most recent forecast is used. Clearly, forecasts and PMU measurements are on completely different time-scales ($\Delta T' \gg \Delta T$), hence the non-triviality of the EnKF. Forecasts are made based on historical data. Previous estimation work based on Kalman filters assumed real-time consumption data. This strong requirement is relaxed with forecasts.

\textbf{Data requirements:} both the SoA and PASE approaches require a forecasting method as well as sample power consumption traces (active and reactive) from the system at the level of each distribution transformer, from which the forecasting method can be calibrated. Using the data, error parameters can be obtained offline. Let $e_i(t)$ be the forecast error at bus $i$ and time $t$ (for active power, for example); $e_i(t)$ is assumed to be a stationary random process. Moreover, forecasts are assumed to be unbiased ($\text{E}[e_i(t)] = 0$) and the variance of the errors ($\text{E}[e_i(t)^2]$) to be known. The estimation of the variance of the forecast errors comes from the acquired data. The assumption of an unbiased forecast is a strong hypothesis, although it is almost always used by researchers \cite{schenato_bayesian_2014}.

The proposed PASE method needs two additional information that can be derived from the same sample data: a load evolution model (which will be discussed in Section \ref{evolution_model}) and the forecast error correlation coefficient, evaluated between two (computation) time-steps at a given bus (i.e., $\text{E}[e_i(t) e_i(t-\Delta T)]$). Given that the data samples are needed for both methods, not much work is involved to derive these additional quantities from it.

Finally, the load forecast errors are assumed to be uncorrelated between buses, an assumption often made in the literature \cite{schenato_bayesian_2014}. 



\textbf{System state:} it is represented by state vectors; different (equivalent) state representations may be used depending on their ease of use in the problem formulation. For example,
$$\mathbf{y}[t] = [\mathbf{V}[t]^T, \mathbf{\delta}[t]^T]^T$$ 
is a possible state vector representation, where $\mathbf{V}[t]$ is the vector of voltage magnitudes at each bus, and $\mathbf{\delta}[t]$ the vector of voltage angles. Another way is to define $\mathbf{x}[t] = [\mathbf{P}[t]^T, \mathbf{Q[t]}^T]^T$ where $\mathbf{P}[t]$ and $\mathbf{Q}[t]$ denotes the vectors of active and reactive power injections at each bus, respectively. Note that the power-flow equations link the state-vectors $\mathbf{x}$ and $\mathbf{y}$. A third way, used in theoretical formulations, is $\mathbf{w}[t] = [ \underline{v}_1[t], \hdots, \underline{v}_{|I|}[t]]^T$ where $\underline{v}_i[t]$ is the voltage phasor at bus $i$, time $t$; this can also similarly be related to other representations.

\textbf{Limitations:} In this work, unbalanced system, distributed generation and biased measurements are not considered and are left for future studies. 


\vspace{-6pt}
\section{State-of-the-Art DSSE Method}
	\label{SoA}

The SoA method \cite{monticelli_state_1999} used to solve the DSSE problem is a snapshot approach and uses a nonlinear WLS objective function. Given the system characterized by the sets $I, B, S$ and the mapping $\mathscr{S}$, the system state, at a given time, is estimated using an overdetermined set of equations. In the following, the time dependency of the variables is dropped for better readability. The variables to be determined are the $2|I|$ state variables. Each measurement adds one constraint. There are either 2 or 4 measurements per bus (active/reactive power forecast, voltage magnitude, and angle), depending on whether there is a PMU at the bus. The number of constraints is given by $M = 2|I|+2|S|$.
The PMU measurements and the forecasts are stored in a vector $\mathbf{z}$ of length $M$, and are related to the system state as per the following model:
    $\mathbf{z} = f(\mathbf{y}) + \eta$
where $f$ is the function that maps the state vector to the measurement vector, and $\eta$ is the vector containing the noise term of each measurement. For example, $f(y) = [\mathbf{V}(y)^T, \delta (y)^T, \mathbf{P}(y)^T, \mathbf{Q}(y)^T]^T$ where $\mathbf{V}(y)$ and $\delta(y)$ are the vectors, respectively, containing the voltage magnitude and angle measurements at the buses with PMUs and $\mathbf{P}(y)$ and $\mathbf{Q}(y)$ are vectors of active and reactive power forecasts of size $|I|$, respectively. Assuming that the measurement errors are uncorrelated and have zero mean, the covariance matrix $\Sigma$ of the error vector $\eta$ is written as, $\Sigma = \text{diag}(\sigma_1^2, ..., \sigma_M^2)$, where $\sigma_m^2$ is the variance of the $m^{th}$ measurement.
The objective function to be minimized at each time-step is given below:
\begin{equation}
    J(\mathbf{y}) = (\mathbf{z} - f(\mathbf{y}))^T \Sigma^{-1} (\mathbf{z} - f(\mathbf{y}))
\end{equation}
Several methods exist to minimize the objective function, the simplest being to iteratively linearize $f$ and solve the resulting objective using the normal equations. 


\vspace{-6pt}
\section{Proposed Method: PASE}
    \label{proposed_solution}
To solve the DSSE problem, PASE, an EnKF-based method, is proposed. Kalman filters are sequential filtering methods. Each iteration is a two steps process: 1) the system state is integrated in time using an evolution model, defining an (\textit{a priori}) state estimate. 2) Available measurements (including pseudo-measurements) are used to correct the estimate and define the updated state. The term \textit{assimilation} is used to refer to the second step. The load evolution model used in this approach is presented in Section \ref{evolution_model}. The idea behind the proposed approach is simple: the additional information provided by the load evolution model and the previously estimated states are used to alleviate the poor quality of pseudo-measurements. 
	
    \vspace{-6pt}
	\subsection{Load Evolution Model}
    	\label{evolution_model}
For each distribution transformer bus, an evolution model for the aggregate load is needed, both for the active and reactive power consumptions. Specifically, the load variation between two (computation) time-steps is considered: let $L_i^p(t)$ and $L_i^q(t)$ denote the instantaneous active and reactive aggregated power respectively, at bus $i$ and time $t$. It is assumed that $L_i^p(t)$ and $L_i^q(t)$ are stationary random processes. The load variation (aka load evolution model) for active and reactive powers are defined as the stationary random processes $L_i^p(t) - L_i^p(t- \Delta T)$ and $L_i^q(t) - L_i^q(t- \Delta T)$ respectively, characterized by their probability density functions (pdf). The mean of the processes is zero and the variance of the processes can be computed from the pdf both for active and reactive powers at bus $i$, denoted $(\sigma^p_i)^2$ and $(\sigma^q_i)^2$, respectively. Such an evolution model is simple and fits within the EnKF framework. The pdf can be derived empirically, for example, from the existing required sample traces, discussed in Section \ref{system_model} as will be explained later.
Clearly a given load evolution model is valid only for systems with similar load compositions, and will vary for different geographical areas. 


	\vspace{-6pt}
	\subsection{Ensemble Kalman Filter}
    	\label{EnKF}
        
The traditional Kalman filter maintains a covariance matrix associated with the state estimate. The EnKF does not use such a matrix and represents the system state pdf using a set of state vectors called ensemble. Such ensemble at time-step $k$ (i.e., time $k \Delta T$) is named $X^k$. The covariance matrix is replaced by the empirical covariance computed from the ensemble. The estimated system state is simply the mean of the ensemble columns. The size of the ensemble, $L$, will impact performance. A small ensemble size will yield faster computations. However the covariance estimate from the ensemble will be less accurate. Therefore there is a trade-off between computational speed and accuracy and a typical choice is a size of $L = 500$ or $1000$. The covariance estimator $\text{cov}(A, B)$ of two ensembles $A, B$ is defined as \cite{evensen_ensemble_2003}:
\begin{equation}
    \text{cov}(A, B) = \frac{1}{N-1}(A - \text{E}[A])(B - \text{E}[B])^T
\end{equation}
where $\text{E}[A]$ is the mean of the column vectors contained in ensemble A. For $\text{cov}(A, A)$ the shorter syntax $\text{cov}(A)$ is used. Each iteration of the EnKF (corresponding to a computation of the state vector at time-step $k$) follows the procedure detailed in Algorithm \ref{alg:EnKF}, each steps of the algorithm are discussed next.

\begin{figure}
\begin{algorithm}[H]
	\caption{Estimation of the state at time-step $k$}
    \label{alg:EnKF}
\begin{algorithmic}[1]
 \renewcommand{\algorithmicrequire}{\textbf{Input:}}
 \renewcommand{\algorithmicensure}{\textbf{Output:}} 
    \REQUIRE $X^{k-1}$, measurements and pseudo-measurements at time-step $k$. 
        \STATE Compute $X_p^k$: integrate the ensemble in time (Eq. \ref{eq:Xp_integration})
        \STATE Compute $X_u^k$: assimilate pseudo-measurements (Eq. \ref{eq:Xu_correlated}) 
        \STATE Compute $X_a^k$: assimilate PMU measurements (Eq. \ref{eq:Xu})
        \STATE $X^k \gets X_u^k$
    \ENSURE Estimated state $\mathbf{\tilde{x}^k} = E[X^k]$ for time-step $k$. 
 \end{algorithmic}
 \end{algorithm}
 \vspace{-15pt}
\end{figure}

	\vspace{-6pt}
	\subsection{Initial Ensemble}
    	\label{initial_ensemble}

The state vector $\mathbf{x} = [\mathbf{P}^T, \mathbf{Q}^T]^T$ (of size $2|I|$) is used. It is chosen given that the load evolution model described in Section \ref{evolution_model} is defined in terms of injected power. The pdf of the state vector $\mathbf{x}$ is represented by an ensemble of size $L$: $X^0 = [\mathbf{x}_1^0, \hdots, \mathbf{x}_L^0]$, $X^0$ is a $2|I| \times L$ matrix containing the ensemble members. The initial ensemble is built by choosing a ``best-guess'' estimate $\mathbf{x}^0$ of the state vector, to which perturbations are added to represent the error statistics of the initial guess. The error distribution chosen for the initial ensemble is discussed in Section \ref{validation}. 

	\vspace{-6pt}
	\subsection{Ensemble Integration}
    

The EnKF is considered at time-step $k$. The prior ensemble $X^k_p$ is obtained by individually integrating forward in time each vector of the ensemble $X^{k-1}$, which was computed at the previous time-step. The integration is such that:
\begin{equation}
    X_p^k = X^{k-1} + [\mathbf{n}_1, \hdots, \mathbf{n}_L] \label{eq:Xp_integration}
\end{equation}  
where $\mathbf{n}_l$ ($l=1, \hdots, L$) are column vectors of size $2|I|$ containing the stochastic noise which accounts for the uncertainties of the load evolution model. Based on the load evolution model defined in Section \ref{evolution_model}, two variance values $(\sigma^p_i)^2$ and $(\sigma^q_i)^2$ are associated to each bus $i$ ($i=1,\hdots, |I|$), respectively for the active and reactive powers. Their values depend on the empirical pdf derived.
Each $n_{i,l}$ and $n_{|I|+i, l}$ ($i=1, \hdots, |I|$) is respectively drawn from a distribution which represents the empirical pdf of the load evolution model. Note that the EnKF can accept any load evolution model.

	\vspace{-6pt}
    \subsection{Assimilation of Pseudo-Measurements}
        \label{time_correlation_pseudo_measurements}
The assimilation of measurements and pseudo-measurements correspond to the update step of the Kalman filter, described at the beginning of Section \ref{proposed_solution}.       

An assumption in Kalman filtering is that the measurement error is white Gaussian noise. Since pseudo-measurements are forecasts and do not depend on the state of the system, they do not satisfy this requirement; instead the forecast error is correlated in time. This problem, which is recurrent in Kalman-based kinematic GPS applications has been solved previously, and a summary of the different existing techniques can be found in \cite{wang_practical_2012}. The solution chosen in this paper is the time-differencing approach described in \cite{petovello_consideration_2009} to remove time-correlated error in the pseudo-measurements. This method was selected for two reasons: 1) it does not require any reinterpretation of the Kalman equations and 2) it does not introduce any latency. 

To remove the correlations, the following process is used. Let the transition matrix $\Psi$ of the time-correlated error be defined as: 
\begin{equation}
	\Psi = \text{diag}(\psi^p_1, \hdots, \psi^p_{|I|}, \psi^q_1, \hdots, \psi^q_{|I|}) \label{eq:psi}
\end{equation}
where $\psi^p_i$ and $\psi^q_i$ ($i=1, \hdots, |I|$) are the forecast error correlation coefficients at bus $i$, respectively for active and reactive powers, introduced in Section \ref{system_model};
 $\Psi$ is diagonal since the forecast errors between buses are assumed to be uncorrelated. $Q$ is defined as the model noise covariance matrix, and is given as:
\begin{equation}
	Q = \text{diag}((\sigma^p_1)^2, \hdots,(\sigma^p_{|I|})^2, (\sigma^q_{|I|+1})^2, \hdots, (\sigma^q_{2|I|})^2 ) \label{eq:Q}
\end{equation}
$R$ is the covariance matrix of the forecast error, of size $2|I| \times 2|I|$. $R$ is diagonal since the forecast errors are assumed not correlated across buses, and is given as:
\begin{equation}
	R = \text{diag}((\sigma_1^{fp})^2, \hdots, (\sigma_{|I|}^{fp})^2, (\sigma_{|I|+1}^{fq})^2, \hdots, (\sigma_{2|I|}^{fq})^2 ) \label{eq:R}
\end{equation}
where $\sigma_i^{fp}$ and $\sigma_i^{fq}$ are the standard deviations of the forecast error at bus $i$, respectively for the active and reactive powers. The pseudo measurements are contained in a vector $\mathbf{d}$ of size $2|I|$. An ensemble $D$ of $L$ perturbed observations is defined such that $D = [\mathbf{d}_1, \hdots, \mathbf{d}_L]$ with each $\mathbf{d}_l = \mathbf{d} + \epsilon_l$ ($l=1, \hdots, L$), where $\epsilon_l$ is a vector drawn from a distribution which models the pseudo-measurement noise. Before establishing the update step, intermediary matrices are defined next, which will be reused for the theoretical derivations. 
\begin{align}
    & H^* = H - \Psi H, ~~~ C = Q H^T \Psi^T, ~~~ D^* = D -\Psi D  \label{eq:Hstar} \\
    & R^* = (R - \Psi R \Psi^T) + \Psi H Q H^T \Psi^T \label{eq:Rstar} 
\end{align}
The updated observation and measurement matrices ($H^*$ and $D^*$, respectively, are computed in (\ref{eq:Hstar}). The updated measurement error matrix $R^*$ is computed in (\ref{eq:Rstar}); $\Psi$ is used to remove the time correlation of the forecast error between two time-steps. The model noise matrix $Q$ is needed to ensure that the noise introduced by the evolution step is retained. Indeed such noise does not have any time correlation component. In this context, the observation matrix $H$ is  the identity matrix (in Section \ref{theoretical_performance} the observation matrix will not be the same). The update equations for the assimilation of pseudo-measurements are given as: 
\begin{align}
    & E = H^* \text{cov}(X^k_p) H^{*T} + R^* + H^* C + C^T H^{*T} \\
    & K = (\text{cov}(X^k_p)H^{*T} + C) E^{-1} \label{eq:K} \\
    & X_u^k = X^k_p + K (D^* - H^* X^k_p) \label{eq:Xu_correlated}
\end{align}

	\vspace{-6pt}
	\subsection{Assimilation of PMU Measurements}
    	\label{PMU_assimilation}

Similar to the pseudo-measurements, the measurements coming from the PMUs are contained in a vector $\mathbf{z}$ of size $2|S|$. An ensemble $Z$ of $L$ perturbed observation vectors is computed such that $Z = [\mathbf{z}_1, \hdots, \mathbf{z}_L]$, with each $\mathbf{z}_l = \mathbf{z} + \xi_l$ ($l=1, \hdots, L$), where $\xi_k$ is a vector drawn from a distribution which models the measurement noise.

The measurements from the PMUs can be related to the state vector using a function $h$, such that $\mathbf{z}_l = h(\mathbf{x}_l) + \gamma_k$, where $\gamma_k$ is an error vector. The function $h(\cdot)$ takes as input the system state and returns a vector containing the measurements that would have been observed considering that particular system state. Given that $\mathbf{x}$ contains the active and reactive powers injected at each bus, $h(\cdot)$ is the power-flow solution; the EnKF does not need to know the analytical expression of $h(\cdot)$. It is the solution given by the LDC's power-flow solver, for example. This makes the EnKF independent of the way power-flows are computed. The cost of such independence is computational: one need to compute $L$ power-flows at each time-step.
Since $h(\cdot)$ is non-linear, the measurements cannot be obtained directly from the state using a simple multiplication by an observation matrix. Instead, $h(\mathbf{x})$ needs to be computed explicitly. A temporary augmented state $\widehat{\mathbf{x}}$ and augmented ensemble $\widehat{X}_u^k$ are used to perform the assimilation, where:
\begin{equation}
	  \widehat{\mathbf{x}}_l = [\mathbf{x_l}^T, h^T(\mathbf{x}_l)]^T, ~~~ 
      \widehat{X}_u^k = [\widehat{\mathbf{x}}_1, \hdots, \widehat{\mathbf{x}}_L]
\end{equation}
The updated ensemble $X_a^k$ is then computed:
\begin{align}
    & X_a^k = X_u^k + K(Z - \widehat{H} \widehat{X}^k_u) \label{eq:Xu}\\
    & K = \text{cov}(X_u^k, \widehat{H} \widehat{X}^k_u) [\text{cov}(\widehat{H} \widehat{X}_u^k) + \text{cov}(Z) ]^{-1}
\end{align}
where $\widehat{H}$ is a selection matrix used to select the rows of the state vector corresponding to the desired measurements.

    \vspace{-6pt}
    \subsection{Theoretical Estimate of Performance}
    	\label{theoretical_performance}
In this section, a method to compute a theoretical estimate of the performance and the improvement achieved by the proposed PASE method is developed. It is based on \cite{schenato_bayesian_2014}, where the authors proposed a technique for estimating a priori the performances of the WLS estimator. Their work is extended in this paper to fit the EnKF and compute the relative gain between the two. The derivation is performed under the following assumptions, also made in \cite{schenato_bayesian_2014}. The state vector is represented by $\mathbf{w} = [\underline{v}_1, \hdots, \underline{v}_{|I|}]^T$. The forecast variance, the forecast error time-correlation and load evolution model variance are assumed to be constant and identical for active and reactive powers. They are denoted respectively $(\sigma^f_i)^2$, $\psi_i^f$ and $(\sigma^d_i)^2$. At each bus $i$, the apparent power magnitude $|S_i^f|$ is used to represent the load forecast. In the analysis framework, the shape of the load evolution model is not need to be known, the value of the variance is sufficient.

The theoretical computations are performed by using a linear Kalman filter. The covariance matrices are made time-invariant in order to obtain a steady-state formulation of the filter \cite{anderson1979optimal}. From this formulation, the covariance matrix of the system state can be computed and used to approximate the performance of the non-linear EnKF. The performance of WLS can also be computed since it can be seen as a Kalman filter that is reset for each new estimation. To evaluate the performance of the two state estimators over a period of time $T$, the average root mean square error of the voltage estimate (ARMSEV) is used as metric:  
\begin{equation}
    \text{ARMSEV} = \sqrt{\frac{1}{T|I|} \sum_{t=0}^{T}\sum_{i=1}^{|I|} \mathbf{E}[|\underline{\hat{v}}_i[t] - \underline{v}_i[t]|^2]}
\end{equation}
where $\underline{v}_i$ is the true voltage at bus $i$ and $\underline{\hat{v}}_i$ the estimated one. A linear version of the power-flow equations is used; it is the first iteration of backward-forward sweep. A vectorized formulation is obtained by using a distribution load flow (DLF) matrix, denoted by $M$, as described in \cite{teng_direct_2003}. The relationship between the injected power at each bus (represented by the vector $\mathbf{s} = [\underline{s}_1, \hdots, \underline{s}_{|I|}]^T$, with $\underline{s}_i$ the injected power at bus $i$) and the state vector is given as:
\begin{equation}
	\mathbf{w} = [V_0, \hdots, V_0] + \frac{1}{V_0}M\times \overline{\mathbf{s}}
\end{equation}
where $\overline{\mathbf{s}}$ is the conjugate of $\mathbf{s}$. Several matrices used by the Kalman equations are defined. The load evolution noise covariance matrix $Q$ expressed in terms of the apparent power, and the forecast error covariance matrix $R_S$ are computed as follows: 
\begin{equation}
	Q = \text{diag}((\sigma^d_1)^2, \hdots, (\sigma^d_{|I|})^2) \label{eq:Qtheoretical}
\end{equation}
\begin{equation}
	R_S = \text{diag}((\sigma^f_1 )^2, \hdots, (\sigma^f_{|I|} )^2) \label{eq:RS}
\end{equation} 
The PMU measurement error covariance matrix is approximated by assuming that the variance of the voltage error when projected onto the real and imaginary axes is the same and equal to $\sigma_{PMU}^2 V_0^2$, where $\sigma_{PMU}^2$ is the relative variance of the PMU measurements such that $R_{PMU} = 2 \sigma_{PMU}^2 V_0^2 \times I_{|S|}$, where $I_{|S|}$ is the $|S|\times |S|$ identity matrix. 

The steady state covariance matrix of the state vector is computed by iterating the Kalman equations. The covariance matrix is denoted by $\Sigma_a^{(\cdot)}$. The iteration number is indicated in the parenthesis $(\cdot)$. Such matrix will converge to a steady state covariance matrix $\Sigma_a^{(ss)}$. For each iteration, two other matrices are used to track the covariance matrix during intermediary steps: $\Sigma_p^{(\cdot)}$ and $\Sigma_u^{(\cdot)}$. They represent respectively the covariance matrix of the prior state and the state after assimilation of PMU measurements. At iteration $0$, the prior covariance matrix of the state is computed such that: 
\begin{align}
    & \Sigma_p^{(0)} = M \times R_S \times M^H \label{eq:priorCovarianceMatrix}
\end{align}
where $(\cdot)^H$ indicates the Hermitian transpose (transpose conjugate operator). The updated covariance matrix obtained after the assimilation of the PMU measurements is then computed:
\begin{align}
    & \Sigma_a^{(0)} = \Sigma_p^{(0)} - K H \Sigma_p^{(0)} \label{eq:sigma_u} \\
    & K = \Sigma_p^{(0)} H^T (H \Sigma_p^{(0)} H^T + R )^{-1} \label{eq:K_sigma_i}
\end{align}
where $H$ is the observation matrix for PMU measurements. It is a selection matrix that relates state variables to the measurement vector. One can estimate the ARMSEV performance of WLS based on $\Sigma_a^{(0)}$: $\text{ARMSEV}_{\text{WLS}} = \sqrt{\frac{1}{|I|}\text{trace}(\Sigma_a^{(0)})} \label{eq:ARMSEV_WLS}$. 

Any iteration $it$ ($it \neq 0$) is performed in 3 steps: first the prior covariance matrix $\Sigma_p^{(it)}$ is computed based on the previous iteration, then the covariance matrix is updated using the PMU measurement covariance matrix, and finally the pseudo-measurements are assimilated. The first two steps are such that (where $H$ is the same matrix as in (\ref{eq:sigma_u})):
\begin{align}
    & \Sigma_p^{(it)} = \Sigma_a^{(it-1)} + M \times Q \times M^H \\
    & \Sigma_{a}^{(it)} = \Sigma_p^{(it)} - K H \Sigma_p^{(it)} \\
    & K = \Sigma_p^{(it)} H^T (H \Sigma_p^{(it)} H^T + R_{PMU} )^{-1} 
\end{align}
The third step differs because of the pseudo-measurement error time correlation (see Section \ref{time_correlation_pseudo_measurements}). The same time-differentiation method is used. The same updated matrices are computed according to (\ref{eq:Hstar})-(\ref{eq:Rstar}) with only a few differences. Now $H$ is the inverse DLF matrix, mapping the state vector to the injected power ($H = M^{-1}$). The forecast error time correlation matrix is such that $\Psi = \text{diag}(\psi_1^f, \hdots, \psi_{|I|}^f)$. Finally, the matrix $R$ used in (\ref{eq:Rstar}) is such that $R = R_S$. The update equations thus become:

\begin{align}
    \Sigma_{a}^{(it)} &= \Sigma_{u}^{(it)} - (\Sigma_{u}^{(it)} (H^*)^{T} + C) \times K^T \\
    K &= \begin{aligned}
         &[\Sigma_{u}^{(it)} * (H^*)^{T} + C] \times [H^* \Sigma_{u}^{(it)} (H^*)^{T} \\
         &  + R^* + H^*C + C^T (H^*){T}]^{-1}
    \end{aligned}
\end{align}

Once the steady state is reached after a few iterations, the theoretical performance of the EnKF can be computed. The ARMSEV error is such that: \newline
$\text{ARMSEV}_{\text{EnKF}} = \sqrt{\frac{1}{|I|}\text{trace}(\Sigma_a^{(ss)})}$. 
The relative gain is expressed as: $\text{Gain}	= \frac{\text{ARMSEV}_{\text{WLS}} - \text{ARMSEV}_{\text{EnKF}}}{\text{ARMSEV}_{\text{WLS}}}$.

\vspace{-6pt}
\section{Validation and Results}
    \label{validation}

The improvement in performance achieved by the proposed PASE method over WLS is evaluated by considering a 33-bus test distribution feeder \cite{baran_network_1989} under normal operations. The WLS estimation problem is modeled in GAMS environment and solved using the MINOS solver. Attention has been paid to avoid potential numerical issues. The ensemble size is set to $L = 500$ and the power flow solutions obtained from $h(\cdot)$ are computed using the backward/forward sweep method \cite{teng_direct_2003}. The system is simulated over a period of 24 hours. For the theoretical estimation, 50 iterations ($(ss) = 50$) are enough to compute the steady-state of the state covariance matrix.

	\subsection{Load Evolution Model}
    	\label{validation_load_evolution_model}
    
The (bus) load evolution model was developed using a fine-grained energy consumption dataset from Ontario, Canada. The dataset used to build the model is described in \cite{ardakanian_markovian_2011} and comprises instantaneous active power consumption data from 20 homes, collected over eight months, with a resolution of 6 seconds. 
The dataset is split randomly into two subsets, one for deriving the characterization (training set), and one for the validation process (testing set). No distinction is made between the size of the houses nor the time of the year. The resulting dataset is a collection of a few thousands of traces. Although 20 homes may seem to be a limited sample size, considering the daily power traces independently allows to have a large number of unique profiles. Moreover the 20 households cover a wide range of living area sizes and energy consumption patterns which increases the trace diversity.

\begin{figure}
    \centering
    \includegraphics[width=0.3\textwidth]{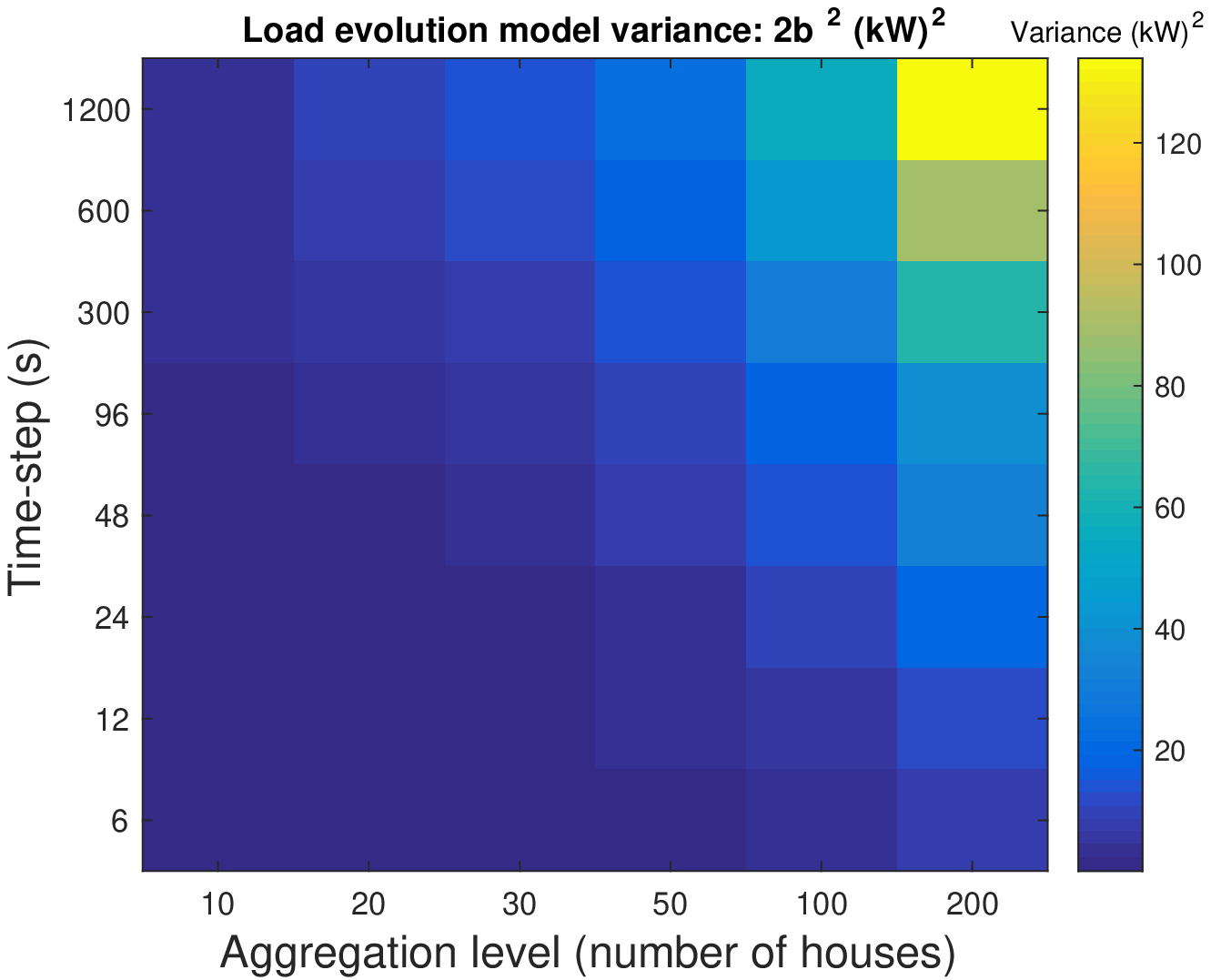}
    \caption{Influence of aggregation level and time step on the scale parameter $b$. The fit of the Laplace distribution is visually good for all the time-steps and aggregation levels considered here.}
    \label{fig:influence_parameter}
\end{figure}

Let $n$ be the number of households connected to a bus. Using the training set, empirical distributions for load changes were constructed for different values of time-steps $\Delta T$ and aggregation levels $n$. A Laplace distribution described by a scale parameter $b$ (and variance $\sigma^2 = 2b^2$) was found to be a good fit. 
The mean value is set to zero since as many positive and negative load changes are expected. This implies that the transition model is the identity, while its uncertainty is characterized by the Laplace distribution. 
The influence of $\Delta T$ and $n$ on the distribution variance is illustrated in Fig.~\ref{fig:influence_parameter}. The variance essentially describes the load variation over time, a small value implying little  variations. It is noted that as $n$ increases and $\Delta T$ shrinks, the value of $\sigma^2$ diminishes. 

It was assumed that load changes are uncorrelated between buses; which can be verified to hold true from the dataset, for any value of $n$ and $\Delta T$ up to 30 minutes. 

The values of $\sigma^2$ are derived empirically as a function of $n$ and $\Delta T$. They are used to compute the evolution step of the EnKF. Since no reactive power consumption dataset was available, a similar model is assumed for reactive power changes. However, active and reactive power consumption changes are assumed to be independent, which is a common assumption in DSSE literature. The proposed method is generic and can be applied to any dataset from across the globe.

    \vspace{-6pt}
    \subsection{Test Distribution System}
    
The 33-bus test feeder data includes active and reactive power loads at each bus; bus-1 is the substation transformer bus, with $V_0$ set to 12.66 kV. The number of houses $n_i$ aggregated at a bus $i$ is selected such that $n_i = n_{11}P_i^{33bus}/P_{11}^{33bus}$ where $n_{11} = 10$ houses and $P_i^{33bus}$ is the static 33-bus active power load at bus $i$.  
The corresponding distribution transformer traces are generated from the second half of the dataset, by summing the desired number of profiles, picked randomly. Each trace is then scaled so that the mean of the profile matches the load values. The values given by the empirical function in Section \ref{validation_load_evolution_model} are scaled accordingly. Because no dataset for reactive power consumption is available, active and reactive power profiles are generated independently from the same dataset.

	\vspace{-6pt}
	\subsection{Measurement Model}
    
The simulation models used for measurements are described in this section. 

\textbf{PMU:} the PMU measurement error is simulated as an additive white Gaussian noise of nominal variance $\sigma^2_{PMU}$, for both voltage magnitudes and angles. The readings $\tilde{V}_s$ and $\tilde{\delta}_s$  provided by the PMU at each bus $s$ ($s \in S \subseteq I$) have an error variance such that $\mathbf{E}[\tilde{a}^2] = \sigma^2_{PMU} * \tilde{a}^2$, where $\tilde{a}$ indicates either the voltage magnitude or angle. The measurement errors are independent across buses, and the voltage magnitude error independent of the angle error. The PMU resolution is set to 1\% ($\sigma_{PMU} = 0.01$); the PMU placement map $\mathscr{S}$ is determined using a greedy method \cite{schenato_bayesian_2014}, i.e., PMUs are sequentially added at the location that provides the most improvement (with 32 load buses, a maximum of 32 PMUs). The placement of PMUs is beyond the scope of this work; many researchers have addressed this issue, see for example \cite{singh_measurement_2009}.

\textbf{Pseudo-measurements}:  the forecasts $P^f_i$ and $Q^f_i$ are taken as the mean value of the load profile generated at each distribution transformer $i$, as in \cite{schenato_bayesian_2014}. They are constant over the simulated period. Using the training set, the nominal standard deviation of the forecast was evaluated and set to $\sigma_0 = 30\%$, for both active and reactive powers, irrespective of the aggregation level. Therefore for each bus $i$, $\sigma_i^{fp} = \sigma_0 P_i^f$ and $\sigma_i^{fq} = \sigma_0 Q_i^f$ (\ref{eq:R}). The constant apparent power forecast $|S_i^f|$ is such that $|S_i^f| = |P_i^f + jQ_i^f|$. Finally each $\sigma^f_i$ (\ref{eq:RS}) is computed as $\sigma^f_i = \sigma_0 |S_i^f|$. Pseudo-measurements with a Gaussian distribution are used as ``best-guess'' initial ensemble (Section \ref{initial_ensemble}).

\textbf{Error time-correlation}: $\psi_i^p$ and $\psi_i^q$ are evaluated as follows: since the same data is used for generating the active and reactive power profiles, $\psi_i^p$ and $\psi_i^q$ are equal. They are evaluated on the training set. Given an aggregation level and a time-step length, load profiles are built. The autocorrelation function $R^e_i$ of the difference between the profile and its mean (representing the forecast error) is computed. The value of $\psi_i^p$ and $\psi_i^q$ is given by $R^e_i(\Delta T)$.



	\vspace{-6pt}
	\subsection{Validation}


\begin{figure*}
    \centering
    \subfloat[][]{\includegraphics[width=0.28\textwidth]{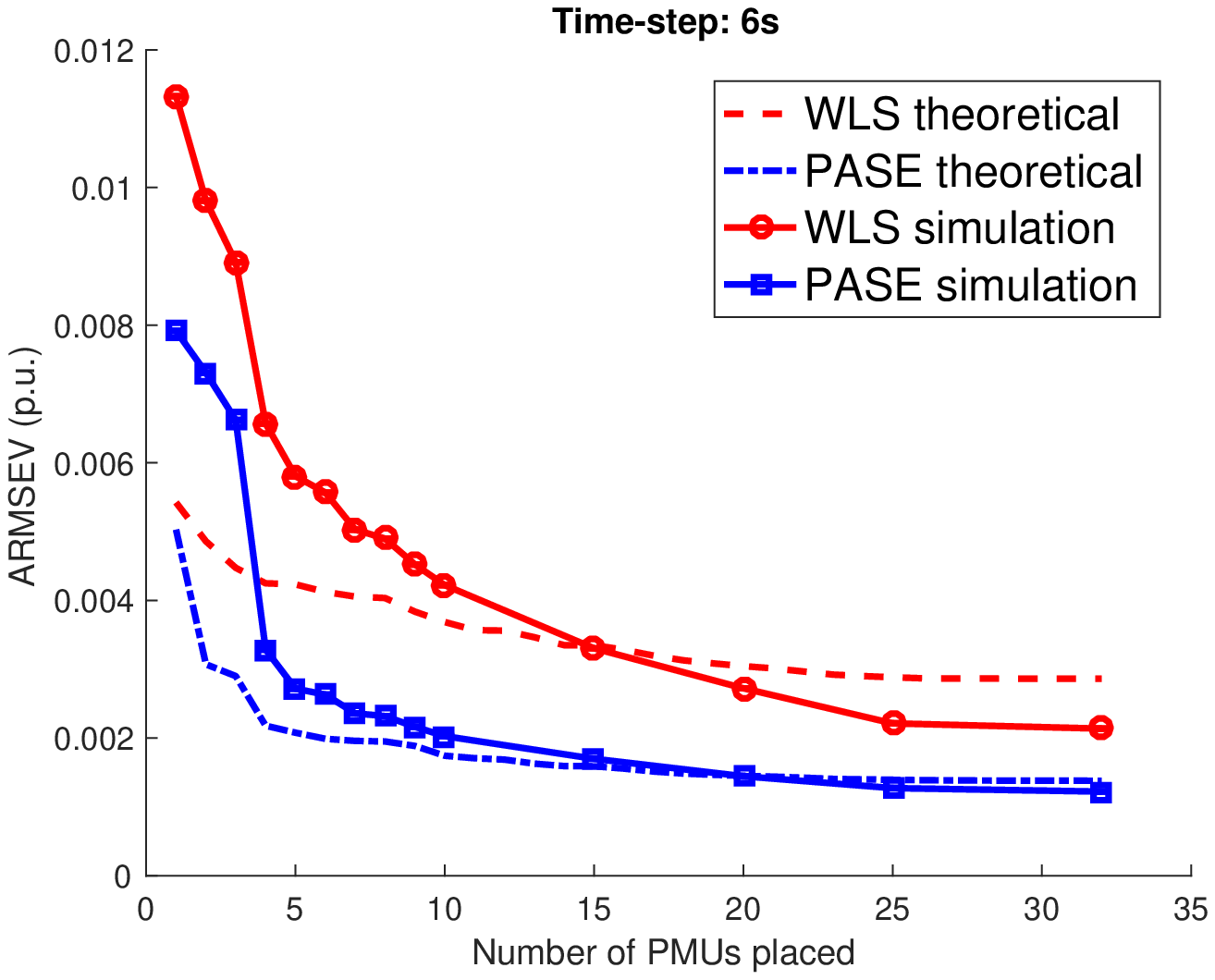}\label{fig:validation_6s_performance}}
    \subfloat[][]{\includegraphics[width=0.28\textwidth]{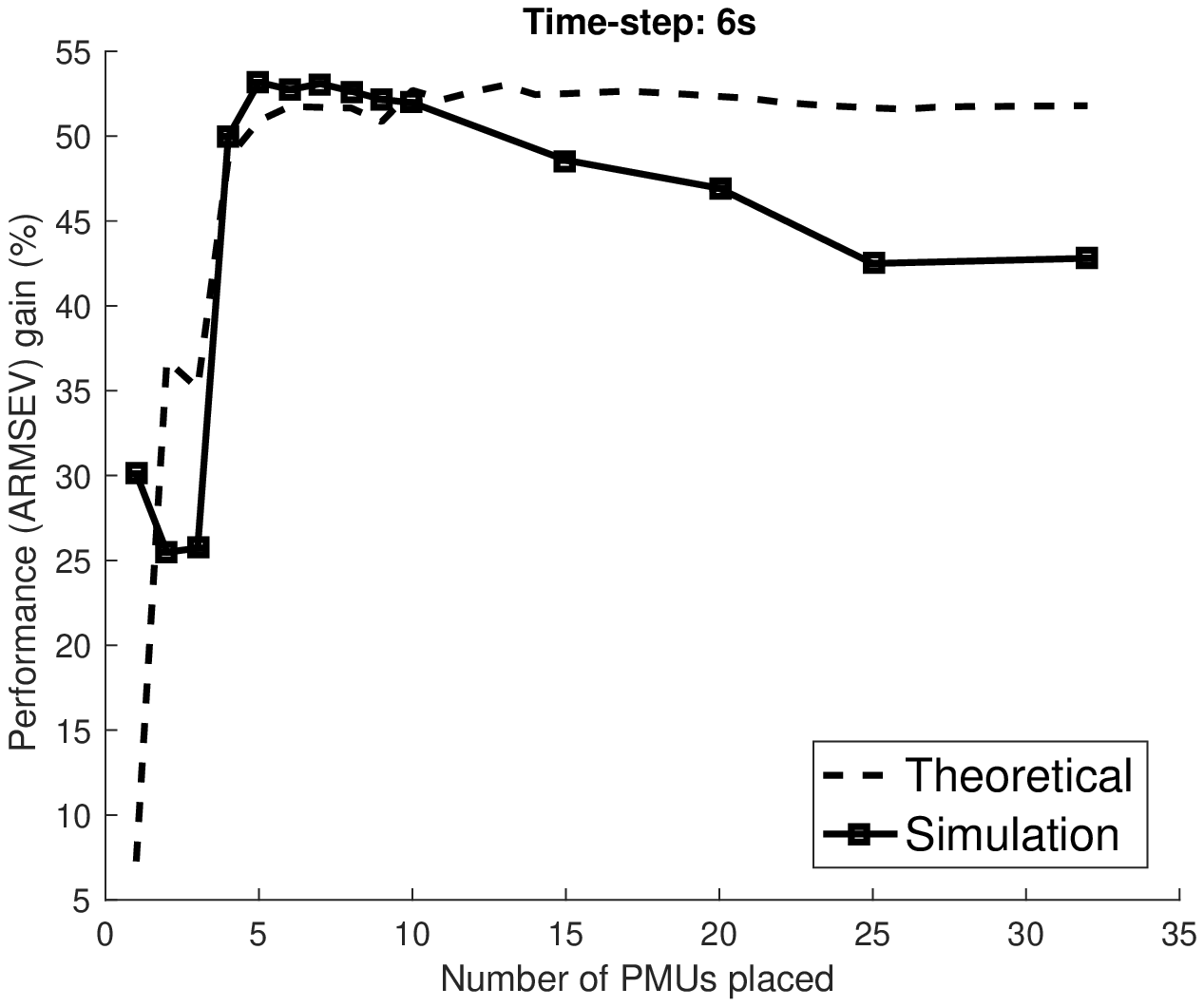}\label{fig:validation_6s_gain}}
    \subfloat[][]{\includegraphics[width=0.28\textwidth]{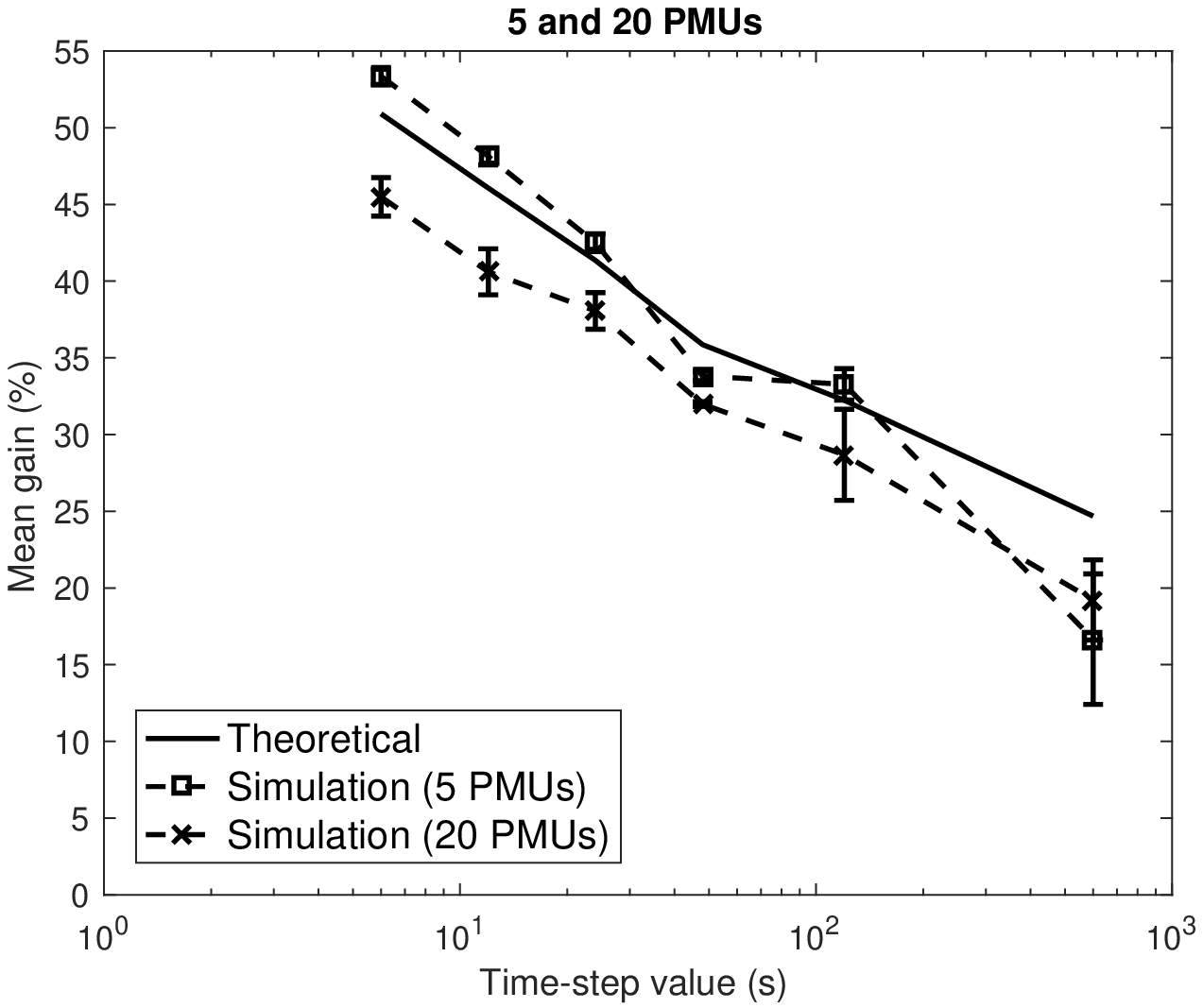}\label{fig:validation_5_20_PMUs}}
    \caption{\protect\subref{fig:validation_6s_performance} ARMSEV value function of the number of PMUs. The performance of the proposed PASE method is compared with WLS. The theoretical results are also compared against simulation results. A lower value means better performance. \protect\subref{fig:validation_6s_gain} Comparison of the gain from using PASE over WLS on ARMSEV depending on the number of PMUs. The theoretical results are compared to the observed gain in simulation. \protect\subref{fig:validation_5_20_PMUs} Influence of the time-step on the mean performance gain. The theoretical results are compared to the observed gain in simulation. The time-step axis has a logarithmic scale. The error bars represent the variance.}
    
\end{figure*}



The theoretical and simulation results are presented in Figs.~\ref{fig:validation_6s_performance}-\ref{fig:validation_5_20_PMUs}, obtained by averaging the results of several realizations. A realization is defined as the observed performance of both the WLS and PASE on the 33-bus system. For each realization, new load profiles are generated based on the testing set, while the other parameters stay the same. The performance of the WLS and PASE are plotted alongside with the theoretical ones in Fig.~\ref{fig:validation_6s_performance}, where a time-step of 6 seconds has been used. WLS has been studied in \cite{schenato_bayesian_2014} using synthetic data. Similar trends are observed here with real data. Note that since WLS is snapshot-based, the size of the time-step does not matter. For PASE, the theoretical results are close to the actual performance observed in simulation as the number of PMUs introduced in the system increases, which validates the theoretical approach. Similar trends are observed for different time-steps. The actual gain brought about by PASE is compared with the theoretical one in Fig.~\ref{fig:validation_6s_gain} for a time step of 6 seconds. 
Finally the influence of the time-step on the gain is compared in Fig.~\ref{fig:validation_5_20_PMUs} for two PMU configurations (5 PMUs and 20 PMUs). Theory and simulation follow the same trend. The gap between theory and simulation is relative to that observed in Fig.~\ref{fig:validation_6s_gain}. For 5 PMUs, a separation between the curves is observed. 

	\vspace{-6pt}
	\subsection{Comparison Between WLS and Proposed PASE Method}   
The results presented in Fig.~\ref{fig:validation_6s_performance} illustrate the improvements achieved by the proposed PASE method. Clearly, using a load evolution model improves the performance of the estimator; given an arbitrary target error of 0.004 p.u., WLS requires more than 10 PMUs while PASE only 4. Even when each bus of the distribution system is monitored by a PMU, the proposed PASE method still brings about an improvement of more than 40\% when using a time-step of 6 seconds. As illustrated in Fig.~\ref{fig:validation_5_20_PMUs}, higher gains are obtained for smaller time-steps. Indeed, for larger time-steps, the load has more chances of changing by a large magnitude between two estimates and thus has less inertia. Even for large time-step (e.g., 10 mins) there is a gain of about 15\%. In practice, the granularity of the time-step depends on the available computational speed. The smallest time-step considered in this work is 6 seconds and represents a lower-bound on what was tried out. In comparison, the DSSE problem was solved at each step in under 1 second.

	\vspace{-6pt}
	\subsection{Engineering Insights}
    
\begin{figure}
    \centering
    \subfloat{\includegraphics[width=0.25\textwidth]{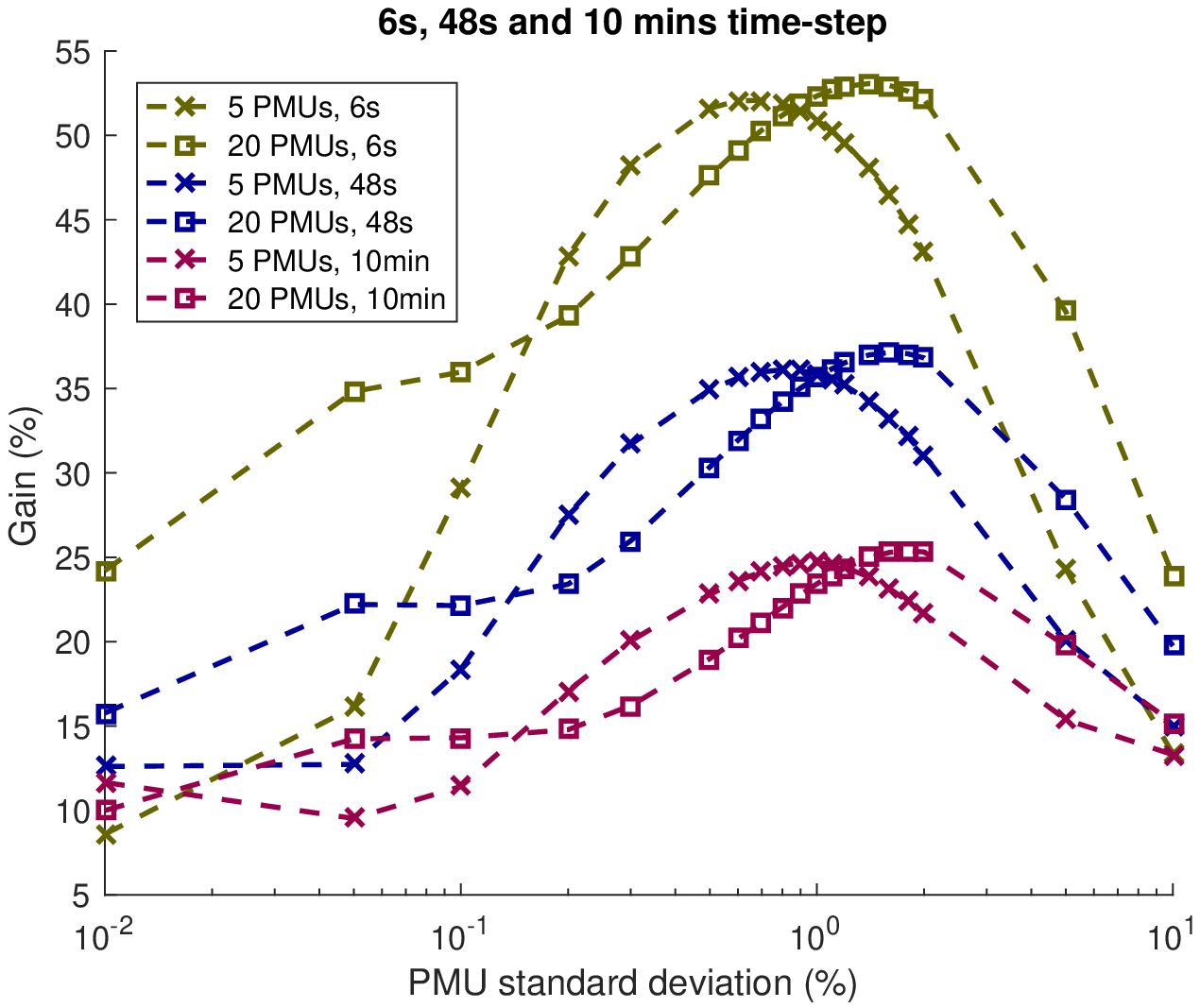}\label{fig:sensor_accuracy}}
    \subfloat{\includegraphics[width=0.25\textwidth]{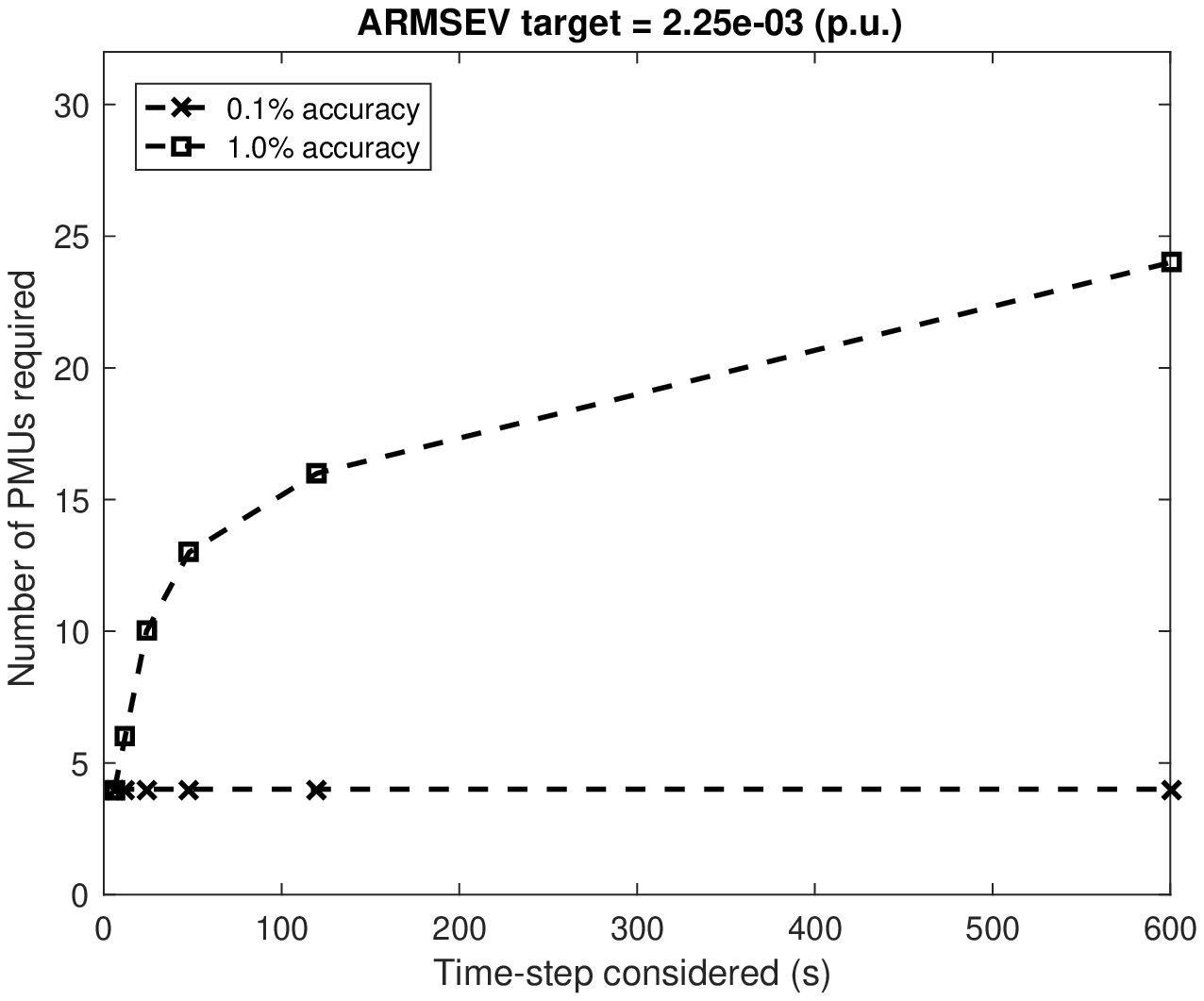}\label{fig:engineering_insight}}
    \caption{\protect\subref{fig:sensor_accuracy} Influence of PMU quality on the (theoretical) gain achieved by PASE over WLS. The PMU accuracy is characterized by its measurement error standard deviation, a lower value means a more accurate PMU. A logarithmic scale is used for the variance axis. \protect\subref{fig:engineering_insight} Minimum number of PMUs required to achieve an average target error of $2.25e-3$ p.u., function of time-step and for two PMU accuracies (computed using the theoretical formulation).}
\end{figure}

    

In practice, the LDC will need to make trade-offs in the choice of the following parameters: number of PMUs, their accuracy and the time scale. The influence of PMU accuracy on the theoretical gain achieved by PASE is shown in Fig.~\ref{fig:sensor_accuracy}, the three parameters considered are depicted in the plot. The maximum gain is attained for a PMU error variance of about 1\%. Clearly as the PMU measurement standard deviation decreases (i.e., the PMU becomes more and more accurate) the gain achieved by PASE decreases since the load evolution model is not as useful in such circumstances. Similarly, as the standard deviation of the PMU increases, the gain decreases, since the load evolution model has to compensate for both poor forecast accuracy and poor PMU measurement accuracy. This figure also illustrates the role of the time-step, the gain achieved by the filtering technique decreasing as the time-step increases, underlining the limits of the load evolution model.

The trade-off between the three parameters considered is illustrated by Fig.~\ref{fig:engineering_insight}: two PMU accuracies are used to draw the plots. An arbitrary target error is fixed and the minimum number of PMUs required is determined as a function of the time-step. Clearly, the time-step has little influence on a very accurate PMU. However, the more accurate the PMU, the more costly it will be. With the same number of PMUs placed in the system (4), choosing a PMU ten times less accurate will provide the same performance given that a time-step small enough (6 seconds) is chosen.

\vspace{-6pt}
\section{Conclusions}
    \label{conclusion}

A novel PASE method for DSSE and its analysis framework were presented. The PASE method performs the fusion of measurements and pseudo-measurements and requires fewer PMUs than WLS to achieve the same estimation error, for time-steps under 15 minutes. Engineering insights were presented highlighting the major trade-offs in the choice of decision variables for the LDC. Using a smaller time-step allows the LDC to relax the requirements on the PMU quality and their number. There are several remaining challenges, such as the influence of distributed generation and its modeling as well as the impact of an unbalanced system on PASE.


\ifCLASSOPTIONcaptionsoff
  \newpage
\fi
\vspace{-6pt}

\bibliographystyle{IEEEtran}
\def\IEEEbibitemsep{0pt plus .5pt}
\bibliography{bib}

\begin{thebibliography}{10}
\providecommand{\url}[1]{#1}
\csname url@samestyle\endcsname
\providecommand{\newblock}{\relax}
\providecommand{\bibinfo}[2]{#2}
\providecommand{\BIBentrySTDinterwordspacing}{\spaceskip=0pt\relax}
\providecommand{\BIBentryALTinterwordstretchfactor}{4}
\providecommand{\BIBentryALTinterwordspacing}{\spaceskip=\fontdimen2\font plus
\BIBentryALTinterwordstretchfactor\fontdimen3\font minus
  \fontdimen4\font\relax}
\providecommand{\BIBforeignlanguage}[2]{{%
\expandafter\ifx\csname l@#1\endcsname\relax
\typeout{** WARNING: IEEEtran.bst: No hyphenation pattern has been}%
\typeout{** loaded for the language `#1'. Using the pattern for}%
\typeout{** the default language instead.}%
\else
\language=\csname l@#1\endcsname
\fi
#2}}
\providecommand{\BIBdecl}{\relax}
\BIBdecl

\bibitem{paudyal_optimal_2011}
S.~Paudyal, C.~A. Canizares, and K.~Bhattacharya, ``Optimal {Operation} of
  {Distribution} {Feeders} in {Smart} {Grids},'' \emph{IEEE Transactions on
  Industrial Electronics}, 2011.

\bibitem{ardakanian_real-time_2014}
O.~Ardakanian, S.~Keshav, and C.~Rosenberg, ``Real-{Time} {Distributed}
  {Control} for {Smart} {Electric} {Vehicle} {Chargers}: {From} a {Static} to a
  {Dynamic} {Study},'' \emph{IEEE Trans. Smart Grid}, 2014.

\bibitem{atanackovic_deployment_2013}
D.~Atanackovic and V.~Dabic, ``Deployment of real-time state estimator and load
  flow in {BC} {Hydro} {DMS} - challenges and opportunities,'' in \emph{2013
  {IEEE} {Power} {Energy} {Society} {General} {Meeting}}, 2013.

\bibitem{monticelli_state_1999}
A.~Monticelli, \emph{\BIBforeignlanguage{en}{State {Estimation} in {Electric}
  {Power} {Systems}}}.\hskip 1em plus 0.5em minus 0.4em\relax Springer, 1999.

\bibitem{von_meier_micro-synchrophasors_2014}
A.~von Meier, D.~Culler, A.~McEachern, and R.~Arghandeh, ``Micro-synchrophasors
  for distribution systems,'' in \emph{{ISGT}, {IEEE} {PES}}, 2014.

\bibitem{rodrigues_low_2016}
R.~N. Rodrigues, J.~K. Zatta, P.~C.~C. Vieira, and L.~C.~M. Schlichting, ``A
  low cost prototype of a {Phasor} {Measurement} {Unit} using {Digital}
  {Signal} {Processor},'' in \emph{{IEEE} {Biennial} {Congress} of
  {Argentina}}, Jun. 2016.

\bibitem{fantin_using_2014}
C.~A. Fantin, M.~R.~C. Castillo, B.~E. B.~d. Carvalho, and J.~B.~A. London,
  ``Using pseudo and virtual measurements in distribution system state
  estimation,'' in \emph{{IEEE} {PES} {T}\&{D}-{LA}}, 2014.

\bibitem{primadianto_review_2016}
A.~Primadianto and C.~N. Lu, ``A {Review} on {Distribution} {System} {State}
  {Estimation},'' \emph{IEEE Transactions on Power Systems}, 2016.

\bibitem{ghosh_distribution_1997}
A.~K. Ghosh, D.~L. Lubkeman, M.~J. Downey, and R.~H. Jones, ``Distribution
  circuit state estimation using a probabilistic approach,'' \emph{IEEE
  Transactions on Power Systems}, 1997.

\bibitem{schenato_bayesian_2014}
L.~Schenato, G.~Barchi, D.~Macii, R.~Arghandeh, K.~Poolla, and A.~V. Meier,
  ``Bayesian linear state estimation using smart meters and {PMUs} measurements
  in distribution grids,'' in \emph{{SmartGridComm}}, 2014.

\bibitem{alam_distribution_2014}
S.~Alam, B.~Natarajan, and A.~Pahwa, ``Distribution {Grid} {State} {Estimation}
  from {Compressed} {Measurements},'' \emph{IEEE Trans. Smart Grid}, 2014.

\bibitem{wang_revised_2004}
H.~Wang and N.~N. Schulz, ``A revised branch current-based distribution system
  state estimation algorithm and meter placement impact,'' \emph{IEEE
  Transactions on Power Systems}, vol.~19, no.~1, pp. 207--213, Feb. 2004.

\bibitem{klauber_distribution_2015}
C.~Klauber and H.~Zhu, ``Distribution system state estimation using
  semidefinite programming,'' in \emph{NAPS}, 2015.

\bibitem{filho_forecasting-aided_2009}
M.~B. D.~C. Filho and J.~C. S.~d. Souza, ``Forecasting-{Aided} {State}
  {Estimation} - {Part} {I}: {Panorama},'' \emph{IEEE Trans. on Power Systems},
  2009.

\bibitem{huang_evaluation_2015}
S.-C. Huang, C.-N. Lu, and Y.-L. Lo, ``Evaluation of {AMI} and {SCADA} {Data}
  {Synergy} for {Distribution} {Feeder} {Modeling},'' \emph{IEEE Trans. Smart
  Grid}, 2015.

\bibitem{sarri_state_2012}
S.~Sarri, M.~Paolone, R.~Cherkaoui, A.~Borghetti, F.~Napolitano, and C.~A.
  Nucci, ``State estimation of active distribution networks: comparison between
  {WLS} and iterated {Kalman}-filter algorithm integrating {PMUs},'' in
  \emph{Innovative {Smart} {Grid} {Technologies} ({ISGT} {Europe})}, 2012.

\bibitem{evensen_ensemble_2003}
G.~Evensen, ``\BIBforeignlanguage{en}{The {Ensemble} {Kalman} {Filter}:
  theoretical formulation and practical implementation},''
  \emph{\BIBforeignlanguage{en}{Ocean Dynamics}}, vol.~53, Nov. 2003.

\bibitem{singh_choice_2009}
R.~Singh, B.~C. Pal, and R.~A. Jabr, ``Choice of estimator for distribution
  system state estimation,'' \emph{IET Gener. Transm. Distrib.}, 2009.

\bibitem{wang_practical_2012}
K.~Wang, Y.~Li, and C.~Rizos, ``Practical {Approaches} to {Kalman} {Filtering}
  with {Time}-{Correlated} {Measurement} {Errors},'' \emph{IEEE Trans. Aerosp.
  Electron. Syst.}, 2012.

\bibitem{petovello_consideration_2009}
M.~G. Petovello, K.~O’Keefe, G.~Lachapelle, and M.~E. Cannon, ``Consideration
  of time-correlated errors in a {Kalman} filter applicable to {GNSS},''
  \emph{Journal of Geodesy}, 2009.

\bibitem{anderson1979optimal}
B.~Anderson and J.~Moore, \emph{Optimal Filtering}.\hskip 1em plus 0.5em minus
  0.4em\relax Prentice-Hall, 1979.

\bibitem{teng_direct_2003}
J.-H. Teng, ``A direct approach for distribution system load flow solutions,''
  \emph{IEEE Transactions on Power Delivery}, Jul. 2003.

\bibitem{baran_network_1989}
M.~E. Baran and F.~F. Wu, ``Network reconfiguration in distribution systems for
  loss reduction and load balancing,'' \emph{IEEE Trans. on Power Del.}, 1989.

\bibitem{ardakanian_markovian_2011}
O.~Ardakanian, S.~Keshav, and C.~Rosenberg, ``Markovian {Models} for {Home}
  {Electricity} {Consumption},'' in \emph{{ACM} {SIGCOMM}}, 2011.

\bibitem{singh_measurement_2009}
R.~Singh, B.~C. Pal, and R.~B. Vinter, ``Measurement {Placement} in
  {Distribution} {System} {State} {Estimation},'' \emph{IEEE Trans. on Power
  Syst.}, 2009.

\end{thebibliography}
%



%





\end{document}